\documentclass{SciPost}

\hypersetup{
    pdfauthor={Florian Herren, Raynette van Tonder},
    pdftitle={On the simulated kinematic distributions of heavy meson decays},
    pdfkeywords={Event Generators, B decays},
    colorlinks,
    linkcolor={red!50!black},
    citecolor={blue!50!black},
    urlcolor={blue!80!black}
}

\usepackage[bitstream-charter]{mathdesign}
\usepackage{algpseudocode}
\usepackage{booktabs}
\DeclareSymbolFont{usualmathcal}{OMS}{cmsy}{m}{n}
\DeclareSymbolFontAlphabet{\mathcal}{usualmathcal}

\fancypagestyle{SPstyle}{
\fancyhf{}
\lhead{\colorbox{scipostblue}{\bf \color{white} ~SciPost Physics }}
\rhead{{\bf \color{scipostdeepblue} ~Submission }}

\fancyfoot[C]{\textbf{\thepage}}
}

\begin{document}

\pagestyle{SPstyle}

\begin{center}{\Large \textbf{\color{scipostdeepblue}{
On the simulated kinematic distributions of heavy meson decays
}}}\end{center}

\begin{center}\textbf{
Florian Herren\textsuperscript{1$\star$}
and
Raynette van Tonder\textsuperscript{2$\dagger$}
}\end{center}

\begin{center}
{\bf 1} Physics Institute, Universit\"at Z\"urich, 8057 Z\"urich, Switzerland
\\
{\bf 2} Institut für Experimentelle Teilchenphysik, Karlsruhe Institute of Technology (KIT), 76131 Karlsruhe, Germany
\\[\baselineskip]
$\star$ \href{mailto:email1}{\small florian.s.herren@gmail.com}\,,\quad
$\dagger$ \href{mailto:email2}{\small raynette.vantonder@kit.edu}
\end{center}

\section*{\texorpdfstring{\color{scipostdeepblue}{Abstract}}{Abstract}}
\textbf{\boldmath{%
Modern measurements in flavour physics rely on accurate simulations of signal and background processes, provided by a wide range of general-purpose and specialised Monte-Carlo event generators. Due to the inclusion of a larger amount of specialised decays of heavy hadrons, EvtGen is often the tool of choice for many scenarios. We investigate the phase-space sampling algorithm of EvtGen and demonstrate that it generates unphysical features in kinematic distributions of semileptonic $B$ decays involving resonances, originating from neglected phase-space factors. We provide a short-term solution to correct the affected simulated samples through reweighting of the hadronic invariant mass distribution.
}}

\vspace{\baselineskip}

%%%%%%%%%% BLOCK: Copyright information
% This block will be filled during the proof stage, and finilized just before publication.
% It exists here only as a placeholder, and should not be modified by authors.
%\noindent\textcolor{white!90!black}{%
%\fbox{\parbox{0.975\linewidth}{%
%\textcolor{white!40!black}{\begin{tabular}{lr}%
%  \begin{minipage}{0.6\textwidth}%
%    {\small Copyright attribution to authors. \newline
%    This work is a submission to SciPost Physics. \newline
%    License information to appear upon publication. \newline
%    Publication information to appear upon publication.}
%  \end{minipage} & \begin{minipage}{0.4\textwidth}
%    {\small Received Date \newline Accepted Date \newline Published Date}%
%  \end{minipage}
%\end{tabular}}
%}}
%}
%%%%%%%%%% BLOCK: Copyright information

\vspace{10pt}
\noindent\rule{\textwidth}{1pt}
\tableofcontents
\noindent\rule{\textwidth}{1pt}
\vspace{10pt}

\section{Introduction}
\label{sec:intro}
Monte-Carlo (MC) event generators play a crucial role in the analysis of data collected at modern high-energy physics experiments \cite{Buckley:2011ms,Campbell:2022qmc}. On the one hand, simulated samples are required for almost every aspect of an experimental analysis, from background suppression and signal extraction to the estimation of efficiencies and unfolding of detector effects. On the other hand, theoretical predictions of observables in complex multi-body processes require the evaluation of high-dimensional phase-space integrals, a task only achievable through Monte-Carlo methods.

The modern event generator ecosystem consists of the general-purpose generators Herwig \cite{Bellm:2025pcw}, Pythia \cite{Bierlich:2022pfr} and Sherpa \cite{Sherpa:2024mfk}, high-energy matrix-element event generators such as \verb|MadGraph5_aMC@NLO| \cite{Alwall:2014hca} and WHIZARD \cite{Kilian:2007gr} as well as specialised generators for QED processes in low-energy $e^+ e^-$ collisions such as BABAYAGA \cite{CarloniCalame:2003yt} and KKMCee \cite{Jadach:2022mbe}. In addition, specialised generators exists for decays of heavy particles, in particular Tauola \cite{Chrzaszcz:2016fte} for $\tau$-lepton decays and EvtGen \cite{Lange:2001uf} for decays of heavy hadrons.

Decays of heavy hadrons are particularly relevant to flavour physics measurements at Belle II, BES III and LHCb, as well as $b$- or $c$-jet tagging at ATLAS and CMS. While all the general purpose generators provide functionality for heavy hadron decays, EvtGen includes a larger amount of specialised decay models, making it the tool of choice in many scenarios.

EvtGen was originally developed by CLEO and BaBar and is part of the main simulation toolchains of ATLAS \cite{ATLAS:2010arf}, Belle II \cite{Kuhr:2018lps,basf2} and LHCb \cite{LHCb:2011dpk}, but is also in use by ALICE, BES III and CMS. Although many aspects of EvtGen have been improved in the last two decades, the core of the event generation algorithm has remained relatively unchanged.

Recently, LHCb found that not all the required phase-space factors are included in EvtGen for the simulation of $B^+\rightarrow K^+ \chi_{c1}(3872)(\rightarrow J/\psi \pi^+ \pi^-)$ decays \cite{LHCb:2022jez}. Accounting for the missing factors, LHCb observed a sizeable $\rho-\omega$ interference contribution in the di-pion invariant mass spectrum, contrary to a previous analysis \cite{LHCb:2015jfc}.

In this work, we will demonstrate that there are more cases in which phase-space factors are not included: semileptonic decays into resonances and nonleptonic decays into two resonances. We derive the probability distribution function describing the EvtGen treatment and study its impact on phenomenologically relevant kinematic distributions.

The remainder of this paper is structured as follows. In Sec.~\ref{sec:sampling} we discuss the phase-space generation within EvtGen in the case of semileptonic decays into resonances and present concrete examples, including potentially affected experimental measurements. In Sec.~\ref{sec::nonlep} we study nonleptonic four-body decays with two intermediate resonances. We present a method to correct existing event samples in Sec.~\ref{sec:reweighting} followed by the conclusion in Sec.~\ref{sec:conclusion}.

\section{Semileptonic decays into resonances in EvtGen}
\label{sec:sampling}
For experimental analyses, the primary event generators only need to produce events with final state, long-lived particles, i.e. stable under the strong interaction, that are passed on to the detector simulation. In the case of semileptonic $B$-meson decays involving resonances, event generators need to simulate a multi-step chain, consisting of a primary $B \rightarrow R \ell \nu$ and subsequent $R\rightarrow P_1 P_2\cdots$ decay. We focus on the case of two final-state hadrons $P_1$ and $P_2$ in the remainder of this section.

For the case of two final-state hadrons, events can be generated through rejection sampling, with the normalised differential decay rate serving as a probability density function (PDF). The normalised $n+2$-fold differential decay rate can schematically be written as
\begin{align}
    \frac{1}{\Gamma}\frac{\mathrm{d}^{n+2}\Gamma}{\mathrm{d}q^2\,\mathrm{d}M^2\,\mathrm{d}\Omega_n} = C\,\left|V_{B\rightarrow R\ell\nu}(q^2, M^2,\Omega_n)\otimes P_R(M^2)\otimes V_{R\rightarrow P_1 P_2}(M^2,\Omega_n)\right|^2~,
    \label{eq::simpledec1}
\end{align}
where $q^2$ is the invariant mass of the lepton-neutrino pair, $\Omega_n$ schematically stands for all relevant angles, $C$ is a normalisation constant, $V_{B\rightarrow R\ell\nu}$ the part of the decay amplitude describing the weak $B\rightarrow R\ell\nu$ transition, $P_R$ is the resonance propagator, assumed to be a dynamic-width Breit-Wigner distribution in EvtGen, and $V_{R\rightarrow P_1 P_2}$ is the decay vertex of the resonance. Each object depends on polarisations of the involved particles and the summation is symbolised by the $\otimes$ symbols.

To simplify this structure, EvtGen assumes that the invariant mass dependence of the decay vertex factorises from the angular structure. Consequently, the angular structure of $P_R$ and $V_{R\rightarrow P_1 P_2}$ can be absorbed into $V_B$, leading to
\begin{align}
    \frac{1}{\Gamma}\frac{\mathrm{d}^{n+2}\Gamma}{\mathrm{d}q^2\,\mathrm{d}M^2\,\mathrm{d}\Omega_n} = C\,\left|\tilde{V}_{B\rightarrow R\ell\nu}(q^2, M^2,\Omega_n)\,\tilde{P}_R(M^2)\,\tilde{V}_{R\rightarrow P_1 P_2}(M^2)\right|^2~.
    \label{eq::simpledec2}
\end{align}
In EvtGen, the resonance propagator is given by
\begin{align}
\tilde{P}_R(M^2) = \frac{1}{M^2 - M_R^2 + iM_R\Gamma_R\,|\tilde{V}_{R\rightarrow P_1 P_2}(M^2)|^2}~,
\end{align}
where $M_R$ is the nominal Breit-Wigner mass of the resonance, $\Gamma_R$ the nominal width and the decay vertex is
\begin{align}
    |\tilde{V}_{R\rightarrow P_1 P_2}(M^2)|^2 = \left(\frac{F^{(L_D)}(|\vec{p}_1|)}{F^{(L_D)}(|\vec{p}'_1|)}\right)^2\left(\frac{|\vec{p}_1|}{|\vec{p}'_1|} \right)^{2L_D + 1}~.
    \label{eq::decvertex}
\end{align}
The $F^{(L)}$ are Blatt-Weisskopf damping factors \cite{VonHippel:1972fg}, $L_D$ is the orbital angular momentum of the decay products and the momenta are defined as
\begin{align}
    &|\vec{p}_1| = \frac{\sqrt{\lambda(M^2,M_1^2,M_2^2)}}{2M}~,\quad |\vec{p}'_1| = \frac{\sqrt{\lambda(M_R^2,M_1^2,M_2^2)}}{2M_R}~,\nonumber\\
    &\lambda(x,y,z) = x^2 + y^2 + z^2 - 2(xy+yz+zx)~.
\end{align}
Here, the $M_i$ are the masses of the final-state particles $P_i$. The primary decay vertex $\tilde{V}_{B\rightarrow R\ell\nu}$ contains process-dependent hadronic form factors, but also phase-space factors that can be deduced solely from the angular momentum $J$ of the resonance:
\begin{align}
    &|\tilde{V}_{B\rightarrow R\ell\nu}|^2 = |\vec{p}_l|^2 |\vec{p}_R|^{\max(2 J - 1,\, 1)} \bar{V}_{B\rightarrow R\ell\nu}~,\nonumber\\
    &|\vec{p}_l| = \frac{\sqrt{\lambda(q^2,m_\ell^2,m_\nu^2)}}{2 \sqrt{q^2}}~,\quad |\vec{p}_R| = \frac{\sqrt{\lambda(M_B^2,q^2,M^2)}}{2 M_B}~.\label{eq::decmoms}
\end{align}
Known features such as the peaking structure of $\tilde{P}_R$ and the factorised momentum dependence of the decay vertices can be used to construct an optimal proposal density for the rejection sampling. To this end, the remainder of the primary vertex, $\bar{V}_{B\rightarrow R\ell\nu}$, needs to be overestimated and samples of $(q^2, M^2)$-pairs can be generated. Optimised methods for phase-space sampling in event generation have a long history and are steadily improved \cite{James:1968gu,Byckling:1969luw,Byckling:1969sx,Kleiss:1985gy,Kleiss:1994qy,Kanaki:2000ey,vanHameren:2002tc,Maltoni:2002qb,Gleisberg:2008fv,vanHameren:2010gg,Horin:2025anc,Frederix:2024uvy,Frederix:2026ejl}, with machine learning techniques playing an increasing role in recent years \cite{Klimek:2018mza,Bothmann:2020ywa,Gao:2020vdv,Gao:2020zvv,Danziger:2021eeg,Butter:2022rso,Verheyen:2022tov,Heimel:2022wyj,Bothmann:2023siu}.
In the following sections, we refer to rejection sampling from the PDF in Eq.~\eqref{eq::simpledec2} as \textit{Full PS}.

In the EvtGen algorithm, only the generation of $M$ is optimised. It is drawn from the PDF
\begin{align}
    \frac{\mathrm{d} \mathcal{P}}{\mathrm{d} M} =  C\,|\tilde{P}_R(M^2)\tilde{V}_{R\rightarrow P_1 P_2}(M^2)|^2~.
    \label{eq::Mdensity}
\end{align}
Values of $q^2$ and the angles are generated by overestimating the entire primary decay vertex $|\tilde{V}_{B\rightarrow R\ell\nu}|^2$ for the proposal density.
However, contrary to standard rejection sampling, EvtGen re-uses the value of $M$ it has drawn from Eq.~\eqref{eq::Mdensity}, until a point for the remaining kinematic is generated and does not fully start from the beginning if the event is rejected. After 10000 failed attempts to generate the remaining kinematics with this fixed value of $M$, EvtGen simply accepts the last point. Consequently, EvtGen does not sample from the PDF in Eq.~\eqref{eq::simpledec2}, but instead from
\begin{align}
    \frac{1}{\Gamma_\text{EvtGen}}\frac{\mathrm{d}^{n+2}\Gamma_\text{EvtGen}}{\mathrm{d}q^2\,\mathrm{d}M^2\,\mathrm{d}\Omega_n} = \frac{C_\text{EvtGen}}{2 M}\frac{\left|\tilde{V}_{B\rightarrow R\ell\nu}(q^2, M^2,\Omega_n)\right|^2}{\int\mathrm{d}t\int\mathrm{d}\Omega' \left|\tilde{V}_{B\rightarrow R\ell\nu}(t, M^2,\Omega')\right|^2} \left|\tilde{P}_R(M^2)\,\tilde{V}_{R\rightarrow P_1 P_2}(M^2)\right|^2~,
    \label{eq::simpledec3}
\end{align}
where we included the Jacobian of $2 M$. We provide a discussion of the corresponding EvtGen code in App.~\ref{app::code}.

This treatment has severe consequences: the phase-space suppression from powers of $|\vec{p}_R|$ in Eq.~\eqref{eq::decmoms} is completely absent, resulting in a non-vanishing $M^2$ spectrum at $M^2 = (M_B - m_\ell)^2$. As a result the hadronic invariant mass spectrum features a significantly enhanced tail, in particular for broad resonances. Due to the distortion of the hadronic invariant mass spectrum, the $q^2$-spectrum is also affected with an excess of generated events at low $q^2$.

In the following, we discuss the aforementioned distortions for the phenomenologically relevant case of charged-current semileptonic $B$-meson decays into four charm resonances, collectively referred to as $D^{\ast\ast}$: the broad $D_0^\ast(2300)$ and the $D_1'(2430)$, as well as the narrow $D_1(2420)$ and $D_2^\ast(2460)$.
While the latter are well established and clearly visible resonances in the $D^{(\ast)}\pi(\pi)$ invariant mass spectra \cite{Belle:2007uwr,BaBar:2008cfy,BaBar:2008dar,Belle:2022yzd}, the former are broad structures, likely associated to two resonance poles \cite{Albaladejo:2016lbb,Du:2017zvv,Meissner:2020khl,Du:2025beb}. However, most experimental studies and event generators treat these broad states as single resonances. The form factors of the four decays are simulated with form factor models developed for the narrow-width limit, such as the update of the Isgur-Scora-Grinstein-Wise model (ISGW2) \cite{Scora:1995ty} or the HQET-based Leibovich-Ligeti-Stewart-Wise (LLSW) model \cite{Leibovich:1997tu,Leibovich:1997em}. In the EvtGen implementation of both models, all resonance-mass dependent terms are evaluated with the invariant mass of the final-state system, not the nominal resonance mass. While the correctness of this approach is unclear, this work does not focus on the physics of hadronic resonances but rather on a purely algorithmic problem.
For first studies of $b\rightarrow c$ transitions beyond the narrow-width limit we refer the reader to Refs.~\cite{Gustafson:2023lrz,Manzari:2024nxr,Papucci:2024qbt}.

\subsection{Broad resonances}
Semileptonic decays into broad resonances are more affected than decays into narrow resonances. This can be easily understood, as the propagator contribution in Eq.~\eqref{eq::simpledec3} is less suppressed near the phase-space boundary for larger widths. For the example of $B\rightarrow D_1'(2430)\mu\nu$ decays, simulated with the ISGW2 model, the distortion of the hadronic invariant mass spectrum is significant.
\begin{figure}[h]
    \centering
    \includegraphics[width=0.8\textwidth]{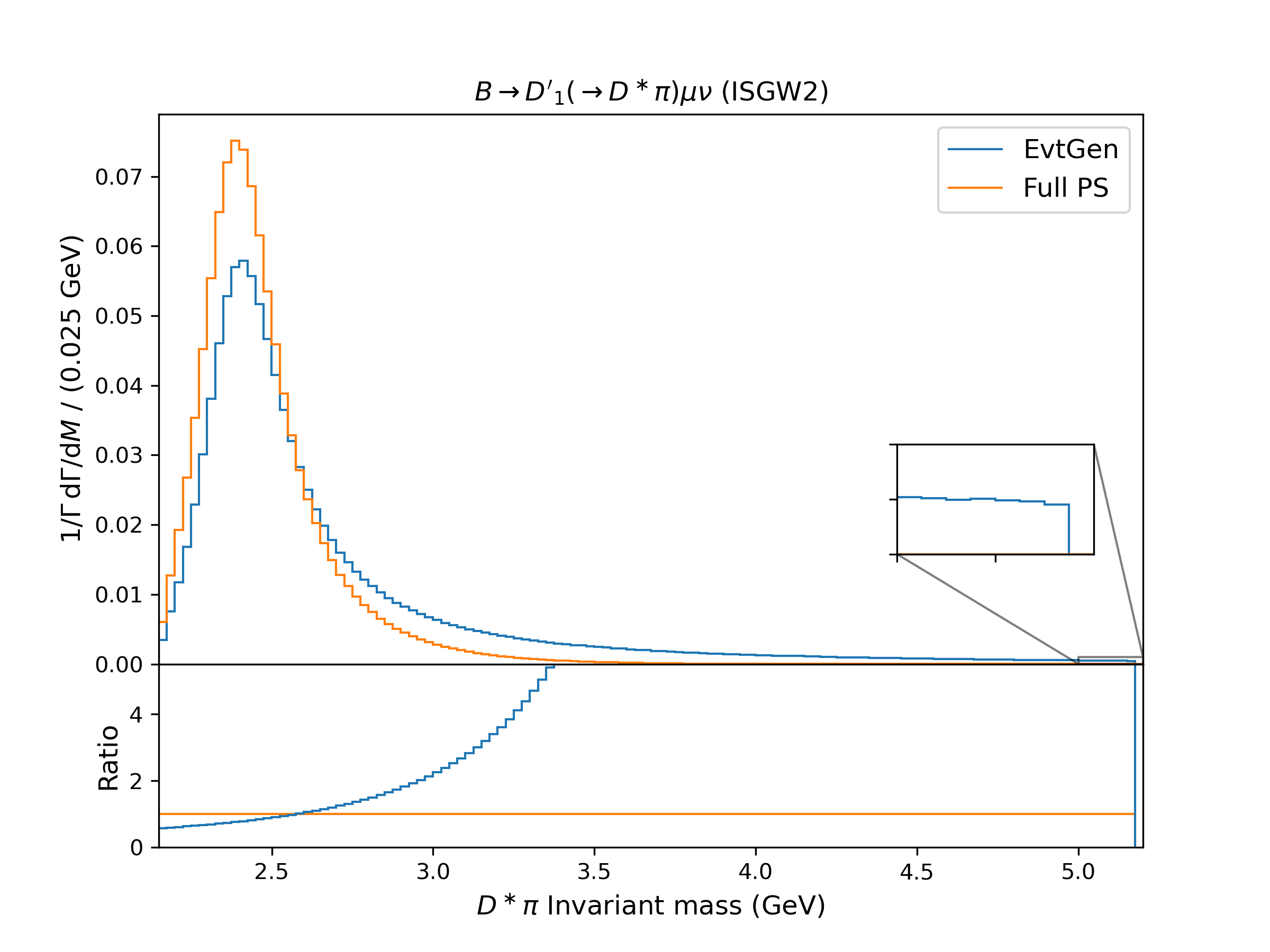}
    \caption{\label{fig:BD1pmunu_M_comp}
        One-dimensional $D^\ast\pi$ invariant mass spectrum of $B^+ \rightarrow D_1'(2430)\mu^+\nu_\mu$ decays generated with EvtGen and by sampling from an implementation of Eq.~\eqref{eq::simpledec2} (Full PS) with the same building blocks as used in EvtGen. The zoom-in focuses on the region near the phase-space boundaries, where the unphysical step is clearly visible. Here and in the following, $10^7$ events are generated for each histogram.
    }
\end{figure}
As can be seen in Fig.~\ref{fig:BD1pmunu_M_comp}, the bulk of the distribution is underestimated due to the long tail and even the peak of the distribution shifts. Furthermore, the distribution does not vanish smoothly at the kinematic boundary $M = M_B - m_\mu$, but exhibits a step.

\begin{figure}[h]
    \centering
    \includegraphics[width=\textwidth]{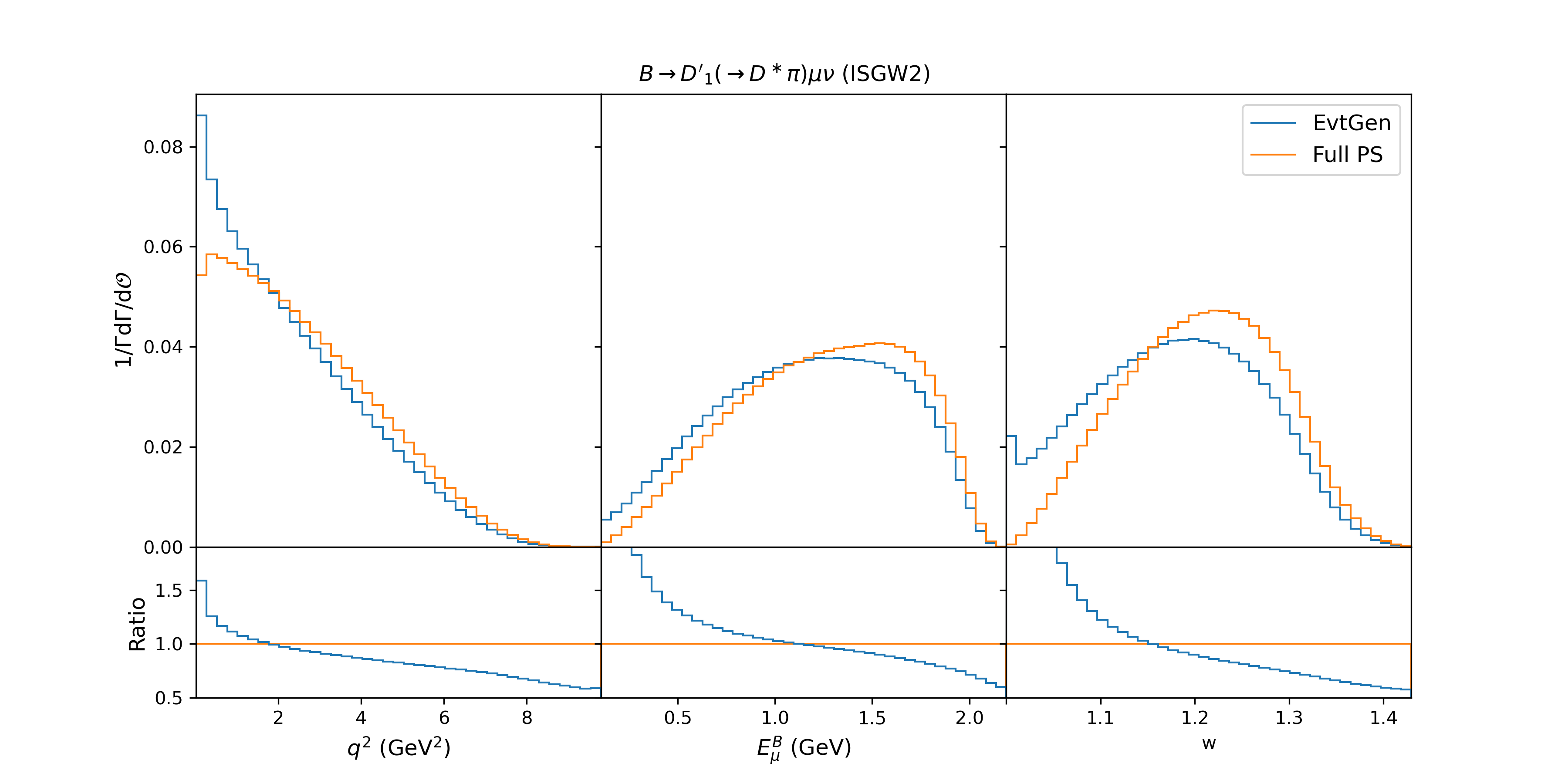}
    \caption{\label{fig:BD1pmunu_q2El_comp}
        One-dimensional $q^2$, $E_\mu^B$ and $w$ spectra of $B^+ \rightarrow D_1'(2430)\mu^+\nu_\mu$ decays generated with EvtGen and by sampling from an implementation of Eq.~\eqref{eq::simpledec2}.
    }
\end{figure}

The distortion in the invariant mass distribution alters all other kinematic distributions of interest. In Fig.~\ref{fig:BD1pmunu_q2El_comp} the distributions in $q^2$, the muon energy in the B-meson restframe, $E_\mu^B$, as well as the recoil, $w = (M_B^2 - M_{D\pi}^2 - q^2) / (2 M_B M_{D\pi})$, are shown.
The shapes of all three distributions are enhanced at lower values and differences can be larger than $50\%$. In addition, the $w$ distribution contains an unphysical spike at $w = 1$, originating from events with large $M_{D\pi}$ and low $q^2$.
In the case of $B\rightarrow D_0^\ast(2300)\mu\nu$ decays, the effects on the kinematic distributions are of similar size to $B\rightarrow D'_1(2430)\mu\nu$ decays.
Since the source of this issue is the phase-space sampling, all form factor models within EvtGen are affected. Comparable distortions are found in samples produced with the open-source \verb|basf2| framework \cite{Kuhr:2018lps,basf2} and the LLSW model.

For decays into $\tau$ leptons, the distortion of the kinematic spectra is even more significant than for light leptons, due to the reduced phase space volume. In this case, the decay rate does not vanish for $q^2 \rightarrow m_\tau^2$, leading to differences of more than $100\%$ for $q^2 < 4\,\text{GeV}^2$ and a shift of the peak position in both distributions.
\begin{figure}[h]
    \centering
    \includegraphics[width=0.8\textwidth]{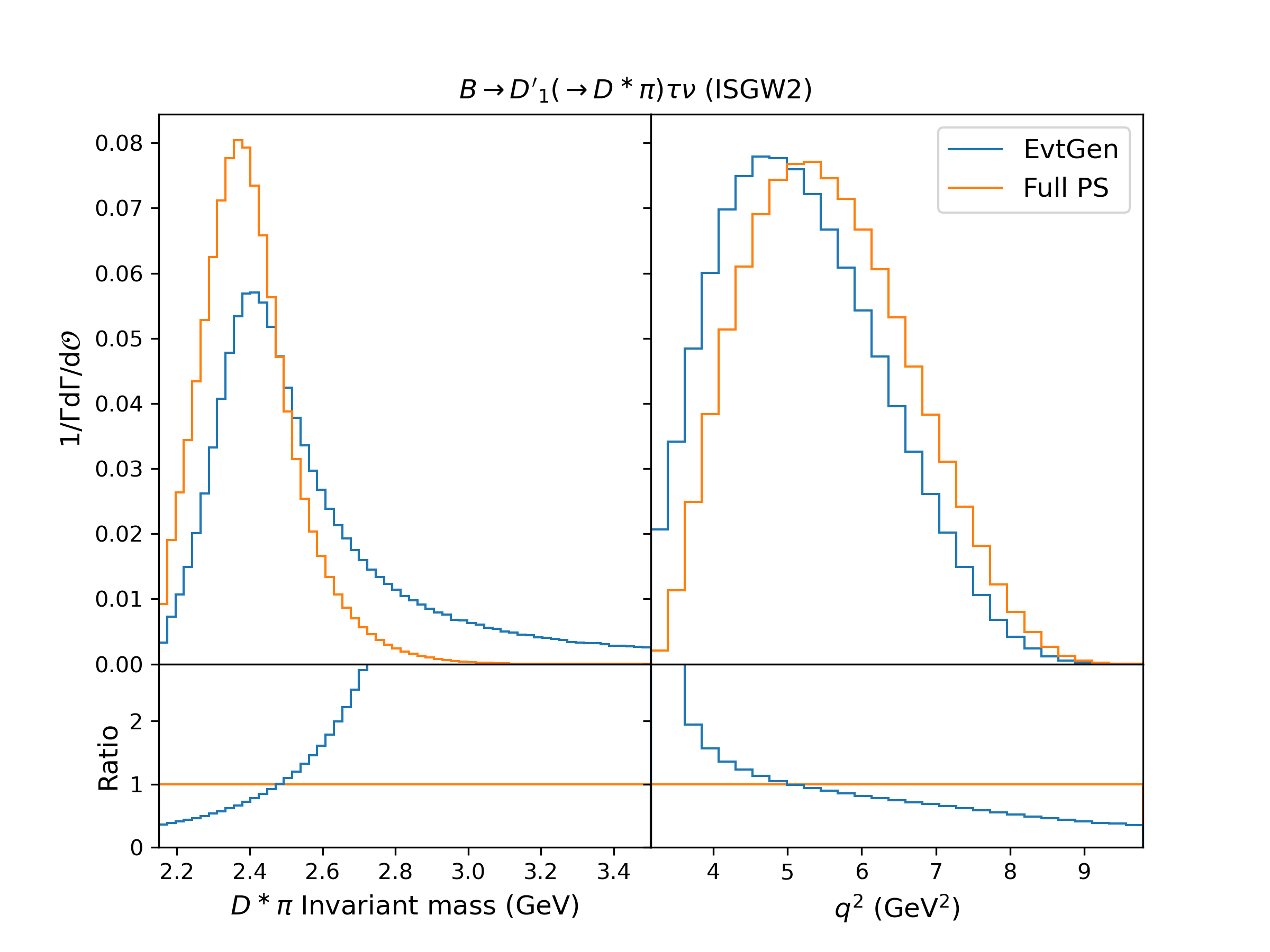}
    \caption{\label{fig:BD1ptaunu_comp}
        One-dimensional $D^\ast\pi$ invariant mass and $q^2$ spectra of $B^+ \rightarrow D_1'(2430)\tau^+\nu_\tau$ decays generated with EvtGen and by sampling from an implementation of Eq.~\eqref{eq::simpledec2}.
    }
\end{figure}

\subsection{Narrow resonances}
In contrast to semileptonic decays into broad resonances, the kinematic distributions of decays into narrow resonances, i.e. the $D_1(2420)$ or $D_2^\ast(2460)$, are significantly less affected.
However, in this case an additional complication arises: EvtGen only simulates events for which
\begin{align}
   M_R - 15\,\Gamma_R < M < M_R + 15\,\Gamma_R~,
\end{align}
where $\Gamma_R$ is the nominal width of the resonance.
This cut affects $2.5\%$ and $0.8\%$ of true $B\rightarrow D_1(2420) \mu\nu$ and $B\rightarrow D_2^\ast(2460) \mu\nu$ events for the LLSW model, respectively. The cut-off leads to enhancements at high values of $q^2$ of about $2\%$ in both cases and a reduction at low $q^2$ of up to $5\%$ at $q^2 \approx m_\mu^2$. Due to the incorrect phase-space sampling, these effects are sub-leading but opposite in sign to the distortions, resulting in only minor differences between the $q^2$-spectra simulated by EvtGen and the distributions obtained from Eq.~\eqref{eq::simpledec2}.

\begin{figure}[h]
    \centering
    \includegraphics[width=0.8\textwidth]{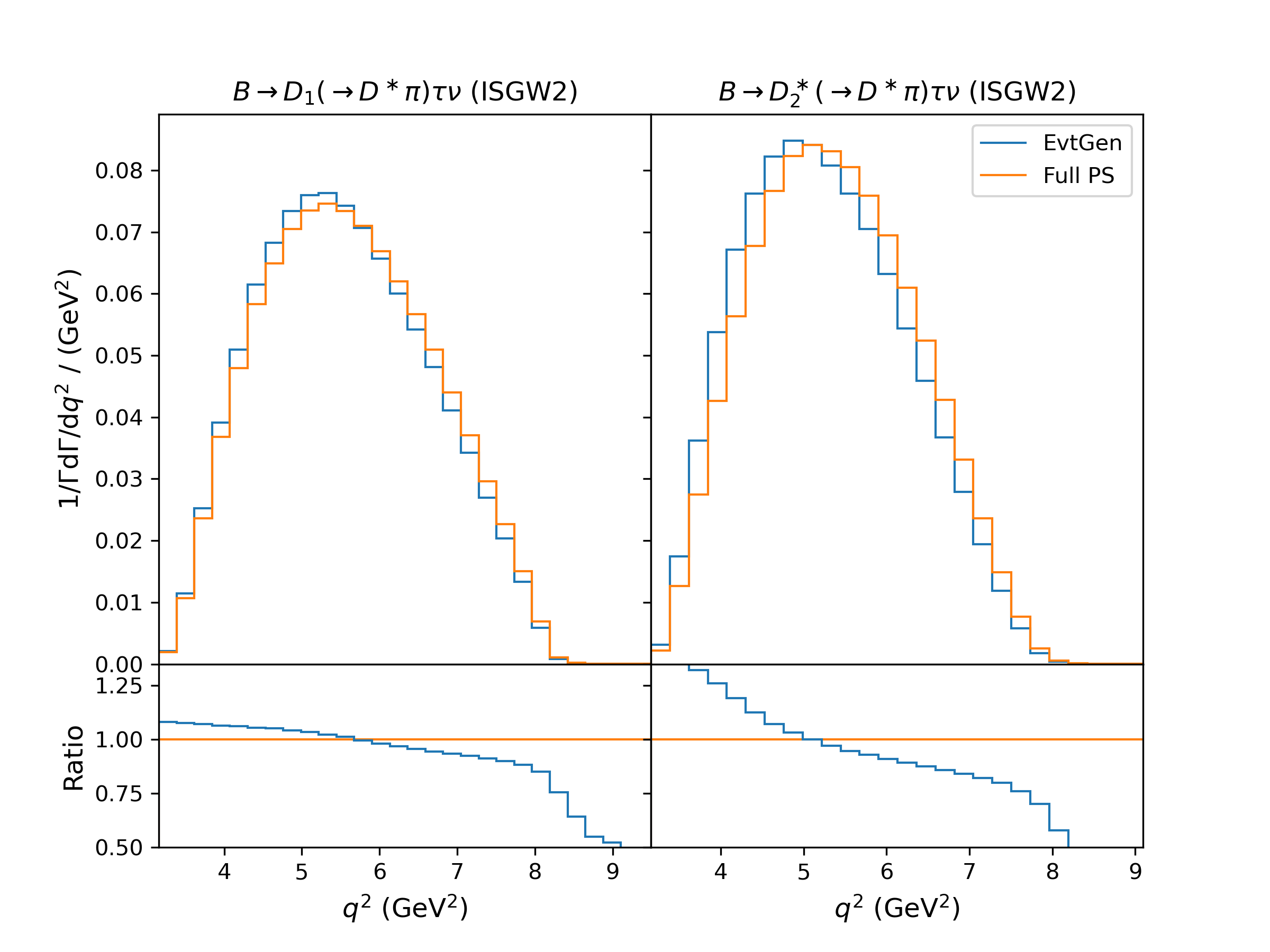}
    \caption{\label{fig::narrow_tau_comp}
        $q^2$ spectra of $B^+ \rightarrow D_1(2420)\tau^+\nu_\tau$ and $B^+ \rightarrow D_2^\ast(2460)\tau^+\nu_\tau$ decays generated with EvtGen and by sampling from an implementation of Eq.~\eqref{eq::simpledec2}.
    }
\end{figure}

Even narrow resonances are non-trivially affected when the available phase-space is restricted. Figure~\ref{fig::narrow_tau_comp} shows the $q^2$-spectra for $B\rightarrow D_1(2420)\tau\nu$ and $B\rightarrow D^\ast_2(2460)\tau\nu$ decays, where we observe similar shape effects compared to the broad $D^{\ast\ast}$ case. These effects are smaller in size, ranging between $10\%-20\%$ at low $q^2$.

\subsection{Possible impact on measurements}
As demonstrated by the LHCb measurement of $\chi_{c1}(3872) \rightarrow J/\Psi\pi^+\pi^-$ decays, an incorrect treatment of the resonance line shapes exerts a significant influence on experimental results and their interpretation. In the following section we examine several experimental analyses that should consider the impact of the underlying mismodelling of resonances and account for this effect in future measurements.

\subsubsection{Hadronic mass moments in \texorpdfstring{$B\rightarrow X_c\ell\nu$}{B → Xclν} decays}
While the distortion of the $q^2$ and $E_\ell^B$ spectra of $B\rightarrow D_0^\ast(2300)\mu\nu$ and $B\rightarrow D_1'(2430)\mu\nu$ decays are significant, kinematic moments of these spectra for inclusive $B\rightarrow X_c\ell\nu$ decays remain mostly unchanged. This can be attributed to the small branching ratios of the broad modes.

The situation for moments of the hadronic invariant mass spectrum is different. They are defined as
\begin{align}
    \left\langle (M^2_X)^n\right\rangle &= \int\mathrm{d}M_X^2\frac{1}{\Gamma}\frac{\mathrm{d}\Gamma}{\mathrm{d}M_X^2}M_X^{2n}\nonumber\\ &= \mathcal{B}(B\rightarrow D\ell\nu)\,(M_D^{2})^n + \mathcal{B}(B\rightarrow D^\ast\ell\nu)\,(M_{D^\ast}^{2})^n + \sum_i  \mathcal{B}(B\rightarrow D_i\ell\nu) \left\langle (M^2_{D_i})^n\right\rangle~,
\end{align}
where the sum over $D_i$ runs over all $D^{\ast\ast}$ and other possible states. Although the branching fraction dictates the relative size of the contribution, the moments of broad states are larger than $(M^2_{D^\ast})^n$. For the case of the first three moments of the $B^+\rightarrow D_1'(2430)\mu\nu$ invariant mass spectrum using the LLSW model we calculate:
\begin{align}
    \left\langle(M^2_{D'_1})\right\rangle_\text{EvtGen} \approx 1.8\times M^2_{D^\ast}~,\quad \left\langle(M^2_{D'_1})^2\right\rangle_\text{EvtGen} \approx 4\times M^4_{D^\ast}~,\quad
    \left\langle(M^2_{D'_1})^3\right\rangle_\text{EvtGen} \approx 11\times M^6_{D^\ast}~.
\end{align}
Consequently, the $B\rightarrow D_1'(2430)\mu\nu$ contribution to the moments is substantially larger than expected from the relative branching fraction of $2.5\%$ \cite{HeavyFlavorAveragingGroupHFLAV:2024ctg}.

To obtain an estimate of the difference between the generator-level moments obtained with EvtGen and from Eq.~\eqref{eq::simpledec2}, we use the branching fractions quoted in the recent Belle II measurement of inclusive $B\rightarrow X_u \ell \nu$ decays \cite{Belle-II:2025pye}, but only take into account the $D$, $D^\ast$ and $D^{\ast\ast}$ contributions.\footnote{There is no evidence of the so-called gap modes $B\rightarrow D^{(\ast)}\eta\ell\nu$ and theoretical studies suggest they are negligible \cite{Gustafson:2023lrz}. }
Since the branching fractions, masses and widths slightly differ between charged and neutral $B$-meson decays, we present the relative shifts
\begin{align}
    \delta_n = \frac{\left\langle (M^2_X)^n\right\rangle_\text{EvtGen}}{\left\langle (M^2_X)^n\right\rangle_\text{Full PS}} - 1
\end{align}
separately in Table~\ref{tbl::MX}.
\begin{table}
\centering
    \begin{tabular}{cccc}
        \toprule
        & $\delta_1$ & $\delta_2$ & $\delta_3$ \\
        \midrule
        $B^+$ & $1.25\%$ & $6.09\%$ & $24.7\%$ \\
        $B^0$ & $1.15\%$ & $5.66\%$ & $23.3\%$ \\
        \bottomrule
    \end{tabular}
    \caption{\label{tbl::MX}
    Relative shift of generator-level hadronic invariant mass moments after correcting for the incorrect phase-space sampling. Estimates are based on masses, widths and branching fractions as used by Belle II in Ref.~\cite{Belle-II:2025pye}.
    }
\end{table}
For all three moments, the generator-level shift is larger than the total systematic uncertainties of the Belle and BaBar measurements \cite{Belle:2006jtu,BaBar:2009zpz} and thus larger than the expected precision from a measurement at Belle II.

The shifts discussed in this section are especially relevant to future analyses that calibrate reconstructed moments to \textit{true} generator-level moments using the method introduced by BaBar \cite{BaBar:2009zpz}, and later adopted for measurements of the $q^2$ moments \cite{Belle:2021idw,Belle-II:2022evt}. Here, a correction factor is obtained by comparing the moments of simulated events with and without selection criteria applied. Events with large hadronic invariant masses are less likely to pass selection criteria, leading to a possible overestimation of correction factors.

\subsubsection{\texorpdfstring{$R(X)$}{R(X)}}
Recently, Belle II reported the first measurement of the ratio of $B\rightarrow X\tau\nu$ w.r.t. $B\rightarrow X\ell\nu$ decays, $R(X)$ \cite{Belle-II:2023aih}. The measurement is systematically limited with the leading sources stemming from exclusive branching fractions and the form factors describing the different modes, as well as an observed mismodelling of the $M_X$ spectrum of $B\rightarrow X_c\ell\nu$ decays. In particular, the analysis reported a deficit at low and an enhancement at high hadronic invariant masses. Both features are mirrored by a deficit at larger values of $q^2$ and an enhancement at lower values, similar to distributions observed by the Belle measurement of the $q^2$ moments \cite{Belle:2021idw} and the Belle II $B\rightarrow X_u\ell\nu$ analysis \cite{Belle-II:2025pye}. While the distortion of the $D_1'(2430)$ and $D_0^\ast(2300)$ line shapes alone can not account for the full effect, additional unknown modes were also simulated through decays of broad resonances to account for the difference between the sum of exclusive modes and the inclusive branching fraction. Consequently, $10\%-20\%$ of all simulated $B\rightarrow X_c\ell\nu$ decays are affected by line shape distortion reported here. The $M_X$ reweighting used in Ref.~\cite{Belle-II:2023aih} leads to a difference between simulated and reweighted lepton energy and $q^2$ distributions similar to the ones shown in Fig.~\ref{fig:BD1pmunu_q2El_comp}. Thus, the mismodelling observed in Refs.~\cite{Belle:2021idw,Belle-II:2023aih,Belle-II:2025pye} could be partially caused by the EvtGen phase-space sampling.

\subsubsection{\texorpdfstring{$R(D^{\ast\ast})$}{R(D**)}}
\begin{figure}[h]
    \centering
    \includegraphics[width=0.8\textwidth]{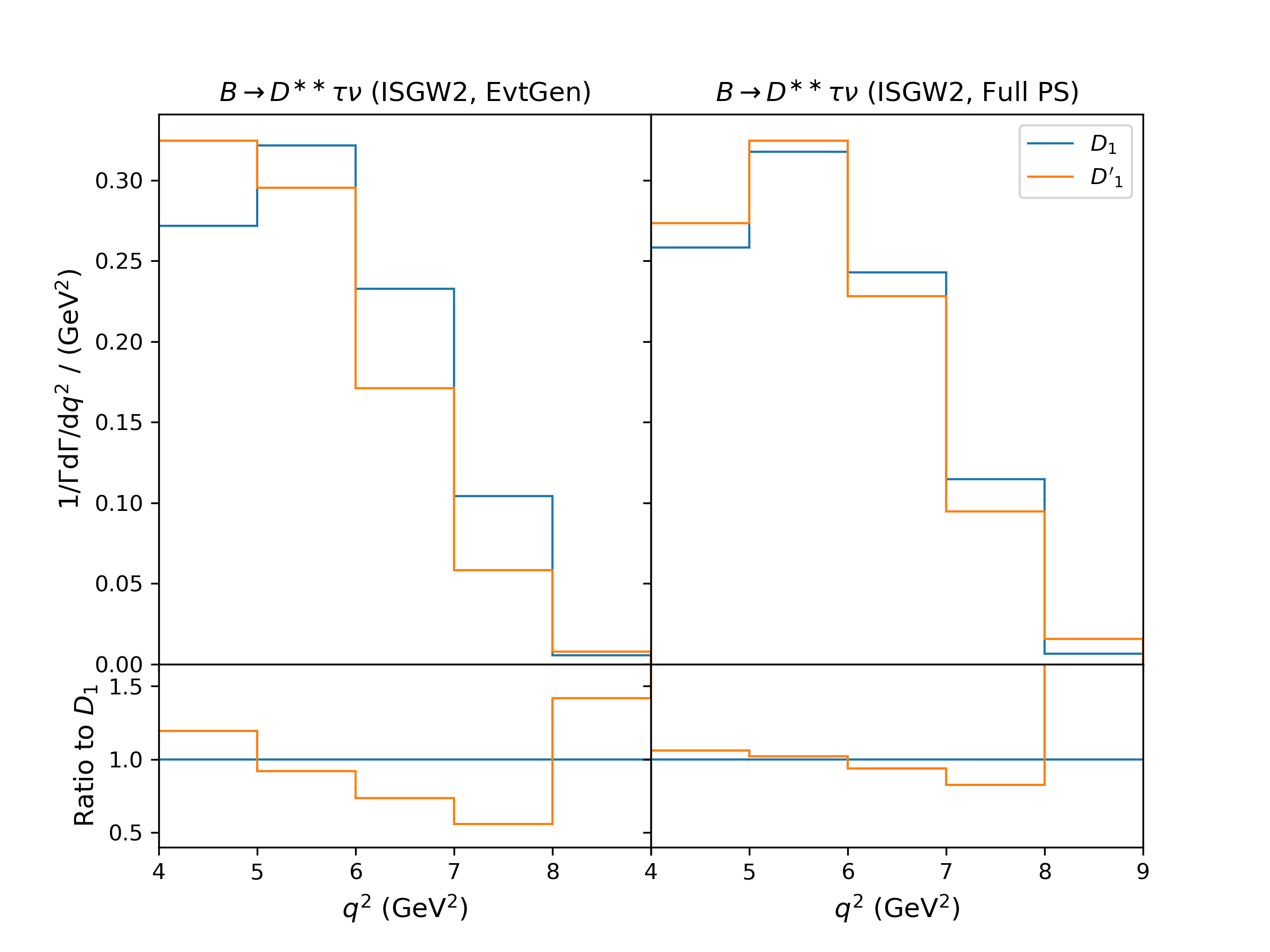}
    \caption{\label{fig::RDstarstar}
        $q^2$ spectra of $B^+ \rightarrow D_1(2420)/D_1'(2430)\tau^+\nu_\tau$ decays generated with EvtGen and by sampling from an implementation of Eq.~\eqref{eq::simpledec2}. The bin size is chosen to match Ref.~\cite{LHCb:2025fri}.
    }
\end{figure}

The first evidence for exclusive $B\rightarrow D^{\ast\ast}(\rightarrow D^\ast\pi)\tau\nu$ decays was recently reported by LHCb \cite{LHCb:2025fri}. In this analysis, the two narrow states were grouped together and constitute the signal component, while the $D_1'(2430)$ mode was treated as background. Two of the three variables in the signal extraction method are directly affected by the distortions discussed in this and the previous section: $\Delta m = M_{D^\ast\pi} - M_{D^\ast}$ and $q^2$. While the impact on the former is negligible for the narrow states, it is sizeable for the $D_1'(2430)$ background, as shown in Fig.~\ref{fig:BD1ptaunu_comp}. However, the signal and background shapes still differ. For $q^2$ the situation is different, since the shapes of all three components change noticeably as demonstrated in Fig.~\ref{fig::RDstarstar}.
Compared to the results generated with EvtGen, directly sampling from Eq.~\eqref{eq::simpledec2} leads kinematic distributions exhibiting similar shapes, in particular of the dominant $D_1(2420)$ and $D_1'(2430)$ components.\footnote{The $D_2^\ast$ component was found to be subdominant in the LHCb analysis and a good description could be achieved without it.} These effects should be studied in ongoing and future analyses as well as taken into account by correcting the kinematic descriptions of the affected decays for meaningful measurements of $B\rightarrow D^{\ast\ast}\tau\nu$ by the LHCb or Belle II experiments.

\section{Distortion of line shapes in nonleptonic decays}
\label{sec::nonlep}
In addition to semileptonic decays, nonleptonic decays into two resonances $R_1$ and $R_2$ are also handled incorrectly by EvtGen. For decays of this type, the correct double-differential decay rate in the two invariant masses can be written in analogy to Eq.~\eqref{eq::simpledec2}:
\begin{align}
    \frac{1}{\Gamma}\frac{\mathrm{d}^{2}\Gamma}{\mathrm{d}M^2\,\mathrm{d}\tilde{M}^2} = C\,\left|\tilde{V}_{B\rightarrow R_1 R_2}(M^2, \tilde{M}^2)\,\tilde{P}_{R_1}(M^2)\,\tilde{V}_{R_1\rightarrow P_1 P_2}(M^2)\,\tilde{P}_{R_2}(\tilde{M}^2)\,\tilde{V}_{R_2\rightarrow P_3 P_4}(\tilde{M}^2)\right|^2~.
    \label{eq::simpledec2_nonlep}
\end{align}
The angular dependence can be neglected in this discussion, since the angular dependence of the primary vertex factorises from the invariant mass dependence for EvtGen models such as the \verb|SVVHELAMP| model.
The primary vertex depends on both invariant masses and thus a completely independent sampling of both variables is not possible.

In EvtGen, each invariant mass is generated assuming that the respective other invariant mass is equal to the nominal Breit-Wigner mass $M_{R_i}$ in the primary vertex. In addition, for invariant masses between the fake phase-space boundary at $(M_B - M_{R_i})$ and the actual phase-space boundary, the primary vertex is simply evaluated with both invariant masses set to the corresponding resonance mass, effectively disabling it.\footnote{Again, the discussion in App.~\ref{app::code} provides references to the relevant code.}
The resulting PDFs for the invariant mass distribution are simply
\begin{align}
        &\frac{\mathrm{d} \mathcal{P}_1}{\mathrm{d} M} =  C_1\,|\hat{V}(M^2)\tilde{P}_{R_1}(M^2)\tilde{V}_{R_1\rightarrow P_1 P_2}(M^2)|^2~,\nonumber\\
         &\frac{\mathrm{d} \mathcal{P}_2}{\mathrm{d} \tilde{M}} =  C_2\,|\bar{V}(\tilde{M}^2)\tilde{P}_{R_2}(\tilde{M}^2)\tilde{V}_{R_2\rightarrow P_1 P_2}(\tilde{M}^2)|^2~,
    \label{eq::Mdensity2}
\end{align}
where the modified primary vertices are given by
\begin{align}
        &\hat{V}(M^2) = (\theta(M_B - M_{R_1} - M)\tilde{V}_{B\rightarrow R_1 R_2}(M^2, M_{R_2}^2)+ \theta(M - M_B + M_{R_1})\tilde{V}_{B\rightarrow R_1 R_2}(M_{R_1}^2,M_{R_2}^2))~,\nonumber\\
        &\bar{V}(\tilde{M}^2) = (\theta(M_B - M_{R_2} - \tilde{M})\tilde{V}_{B\rightarrow R_1 R_2}(\tilde{M}^2, M_{R_2}^2) + \theta(\tilde{M} - M_B + M_{R_2})\tilde{V}_{B\rightarrow R_1 R_2}(M_{R_1}^2,M_{R_2}^2))~.
    \label{eq::Mdensity3}
\end{align}
The treatment of the primary vertex clearly generates unphysical steps in the invariant mass spectra. Combining Eqs.\eqref{eq::Mdensity2} and \eqref{eq::Mdensity3}, we obtain
\begin{align}
    \frac{1}{\Gamma_\text{EvtGen}}\frac{\mathrm{d}^{2}\Gamma_\text{EvtGen}}{\mathrm{d}M^2\,\mathrm{d}\tilde{M}^2} = \frac{C_\text{EvtGen}}{4M\tilde{M}}\,\left|\hat{V}(M^2)\,\bar{V}(\tilde{M}^2)\,\tilde{P}_{R_1}(M^2)\,\tilde{V}_{R_1\rightarrow P_1 P_2}(M^2)\,\tilde{P}_{R_2}(\tilde{M}^2)\,\tilde{V}_{R_2\rightarrow P_3 P_4}(\tilde{M}^2)\right|^2~.
    \label{eq::evtgen_nonlep}
\end{align}
If the sum of the two invariant masses is larger than the $B$-meson mass, the event is rejected and a new pair is generated. For $B$-meson decays into two light resonances, such as $\rho$ or $K^\ast$, this procedure, while formally incorrect, is largely unproblematic, since the fake phase-space cut-off is already beyond the width cut-off discussed in the previous section. However, in the case of two heavy resonances or for $D$-meson decays, the line shapes are affected in a non-trivial manner.

\begin{figure}[t]
    \centering
    \includegraphics[width=0.8\textwidth]{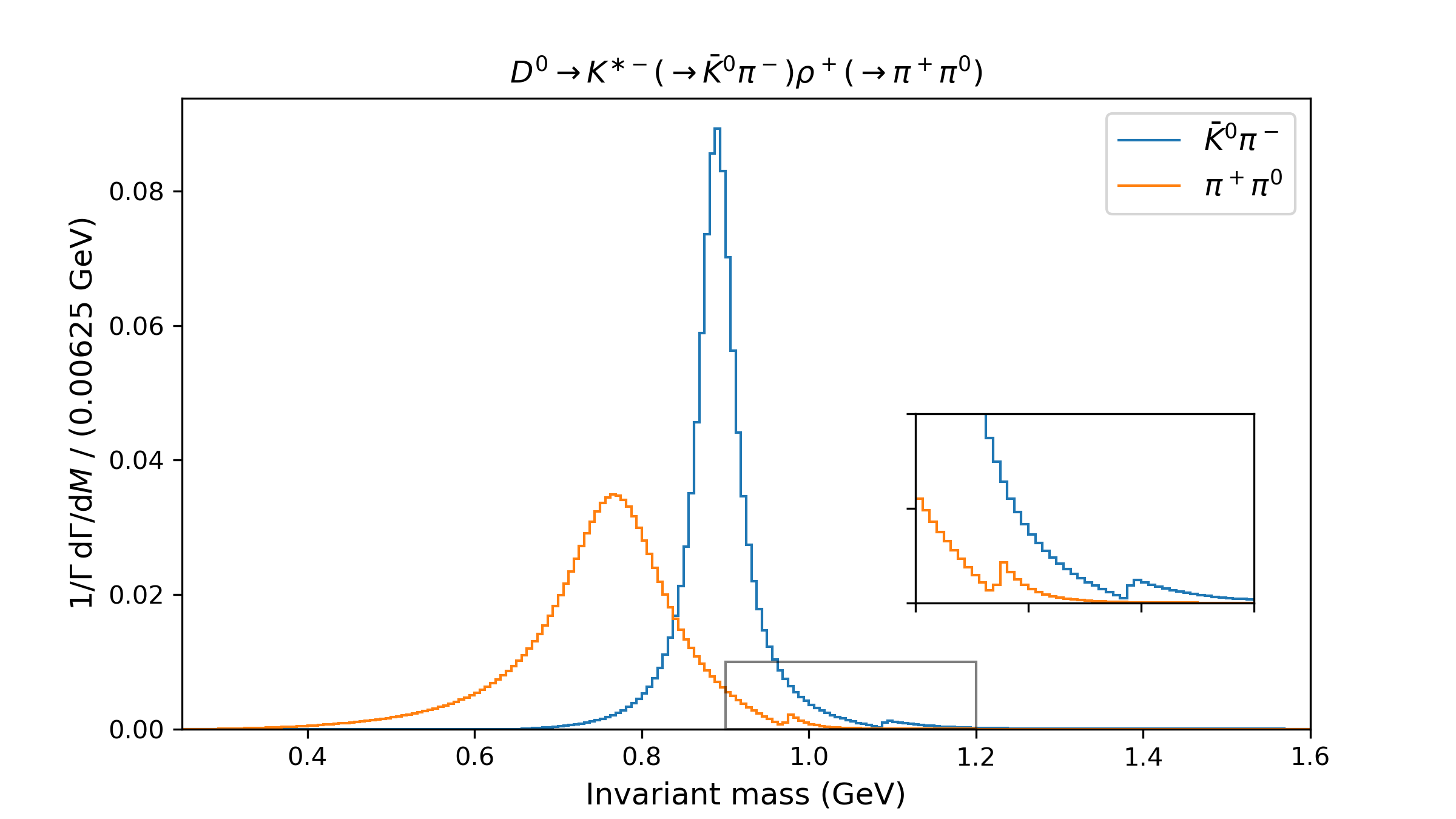}
    \caption{\label{fig:DKstarRho_evtgen}
        One-dimensional $K\pi$ and $\pi\pi$ invariant mass spectra of $D^0 \rightarrow \bar{K}^{\ast -}\rho^+$ decays generated with EvtGen. The zoom-in focuses on the region near the fake phase-space boundaries, where the unphysical steps are clearly visible.
    }
\end{figure}

\begin{figure}[h]
    \centering
    \includegraphics[width=0.8\textwidth]{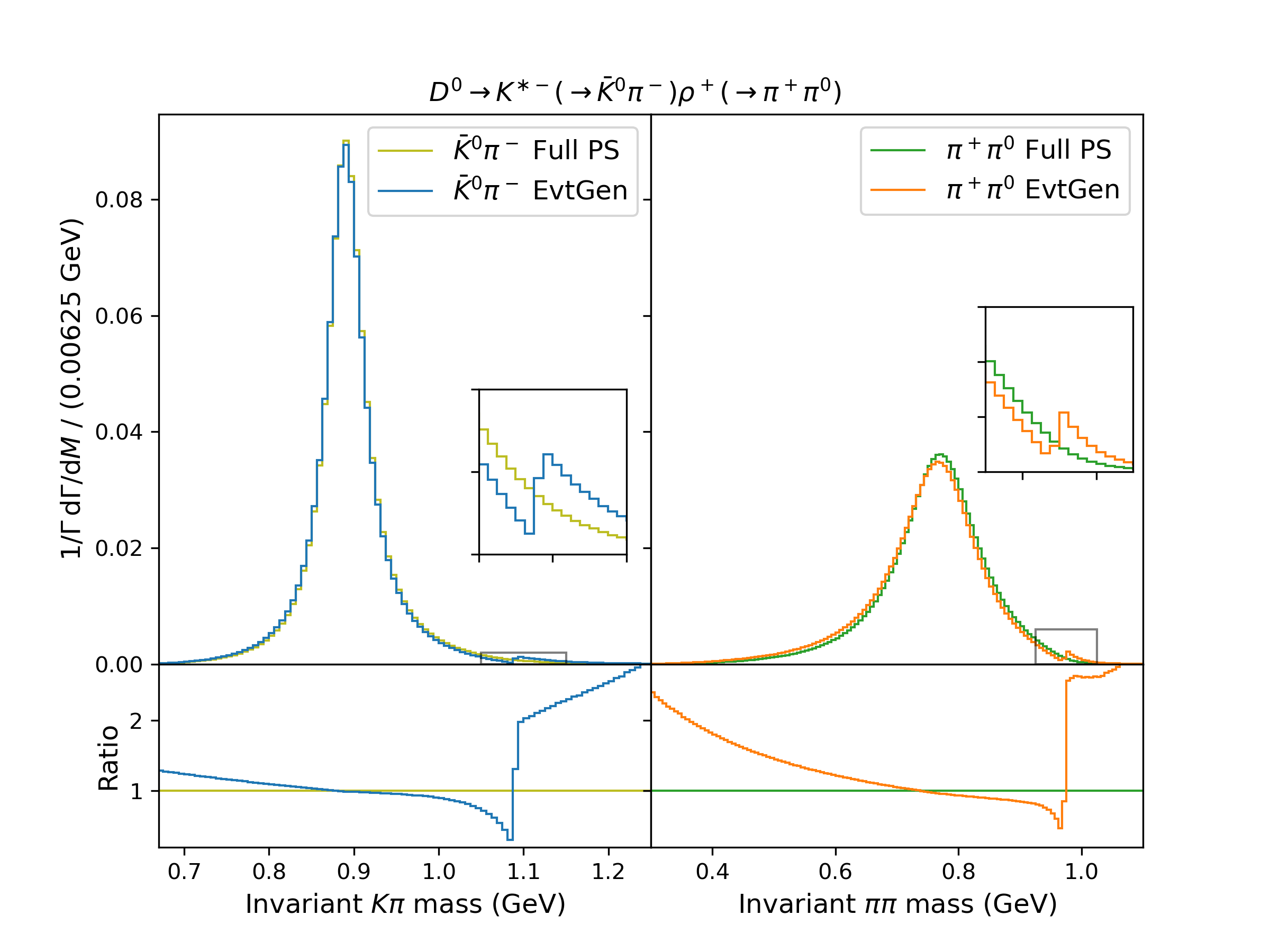}
    \caption{\label{fig:DKstarRho_comp}
        One-dimensional $K\pi$ and $\pi\pi$ invariant mass spectra of $D^0 \rightarrow \bar{K}^{\ast -}\rho^+$ decays generated with EvtGen and based on Eq.~\eqref{eq::simpledec2_nonlep}. The zoom-in focuses on the region near the fake phase-space boundaries and the lower panel shows the ratios of the EvtGen spectra to those from Eq.~\eqref{eq::simpledec2_nonlep}.
    }
\end{figure}

As a concrete example, we consider $D^0 \rightarrow \bar{K}^{\ast -}(\rightarrow \bar{K}^0 \pi^-)\rho^+(\rightarrow \pi^+ \pi^0)$ decays, where the procedure leads to apparent phase-space cut-offs of the invariant mass spectra at $M_D - M^\text{nom}_\rho$ and $M_D - M^\text{nom}_{K^\ast}$, instead of the correct phase-space boundaries $M_D - 2 M_\pi$ and $M_D - M_K - M_\pi$, respectively.
The situation is shown in Fig. \ref{fig:DKstarRho_evtgen}, where the two invariant mass spectra as generated by EvtGen are depicted.
A comparison of the one-dimensional invariant mass spectra from EvtGen and from Eq.~\eqref{eq::simpledec2_nonlep} is shown in Fig.~\ref{fig:DKstarRho_comp}.
In addition to significant differences in the region near the fake phase-space boundary, as well as above, deviations are large at low invariant masses, leading to visible differences in the $\pi^+\pi^0$ invariant mass spectrum. Due to the smaller width of the $K^\ast$, the $\bar{K}^0\pi^-$ invariant mass spectrum is less affected.

\section{Correcting kinematic distributions through reweighting}
\label{sec:reweighting}
Correcting the issues reported in the previous section within EvtGen will require time and careful validation. Additional time will be required for experimental collaborations to incorporate an updated official version of EvtGen into the core collaboration software and produce new MC samples. A short-term method to correct existing event samples for this issue is required.

Fortunately, this is possible by calculating event-wise correction weights from the normalised differential decay rates of the correct and EvtGen distributions. These event-wise weights are calculated according to:
\begin{equation}
    w(\vec{v}) = \frac{\Gamma_{\text{EvtGen}}}{\Gamma}\frac{d\Gamma (\vec{v})/dv}{d\Gamma (\vec{v})_{\text{EvtGen}}/dv}
\end{equation}
Here, the weight $w$ is a function of a set of kinematic variables $\vec{v}$, while the differential decays rates are either determined from analytical expressions or from distributions of generated events.

\begin{figure}[h]
    \centering
    \includegraphics[width=0.8\textwidth]{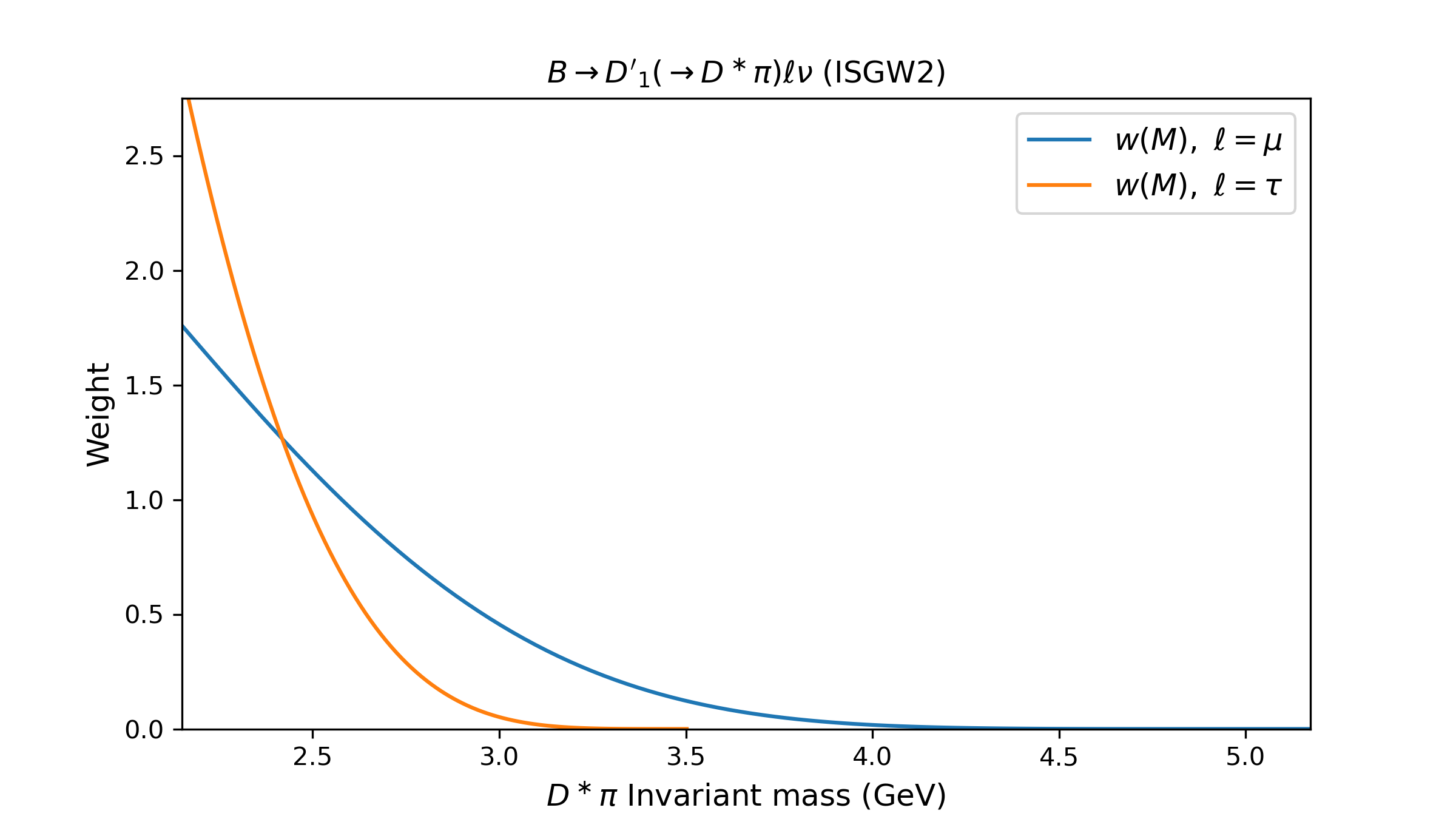}
    \caption{\label{fig:BD1pmunu_weights}
        Weight functions for $B^+ \rightarrow D_1'(2430)\mu^+\nu_\mu$ and $B^+ \rightarrow D_1'(2430)\tau^+\nu_\tau$ in the ISGW2 model.
    }
\end{figure}

For the case of semileptonic decays discussed in Sec.~\ref{sec:sampling} the weights are particularly simple and only a one-dimensional reweighting in $\vec{v} = \{M^2\}$ is required. Following Eq.~\eqref{eq::simpledec3} we obtain the correction weight
\begin{align}
    w(M^2) = 2 M\,\frac{\Gamma_{\text{EvtGen}}}{\Gamma}\label{eq::weightSL}\int\mathrm{d}q^2\int\mathrm{d}\Omega\,\left|\tilde{V}_{B\rightarrow R\ell\nu}(q^2, M^2,\Omega)\right|^2~.
\end{align}
The integrand depends on the form-factor model used to generate the events in EvtGen and the final $q^2$-integration needs to be performed numerically. Two example weight functions for $B^+ \rightarrow D_1'(2430)(\rightarrow D^{\ast -}\pi^+)\ell^+\nu_\ell$ decays in the ISGW2 model are shown in Fig.~\ref{fig:BD1pmunu_weights}.

\begin{figure}[h]
    \centering
    \includegraphics[width=0.8\textwidth]{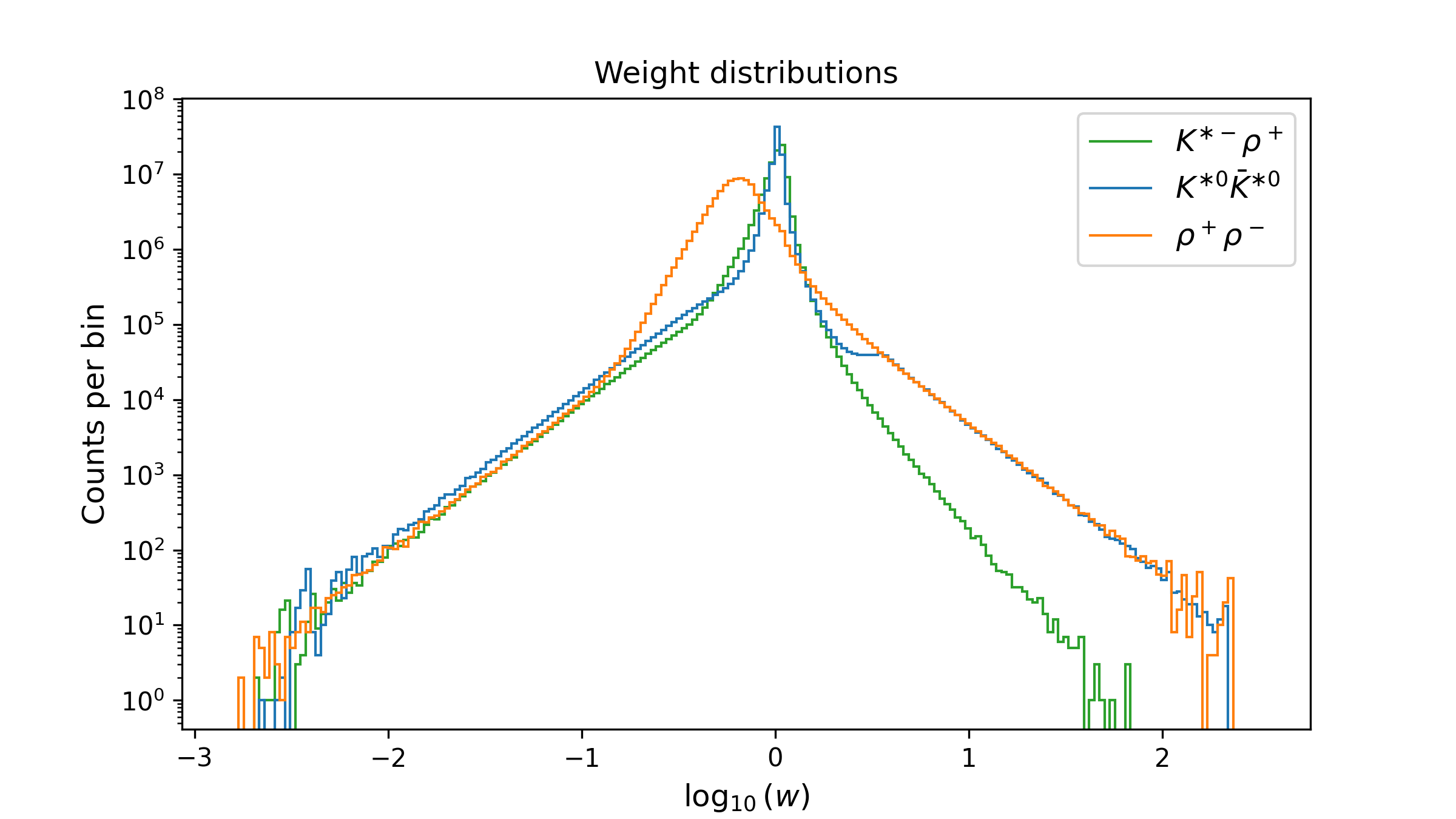}
    \caption{\label{fig:DKstarRho_weights}
        Weight distributions for $10^8$ $D^0\rightarrow K^{\ast -}\rho^+$, $D^0\rightarrow K^{\ast 0}\bar{K}^{\ast 0}$ and $D^0\rightarrow \rho^+\rho^-$ decays.
    }
\end{figure}

The nonleptonic two-resonance decays discussed in Sec.~\ref{sec::nonlep} are slightly more complicated. Here, a two-dimensional reweighting with $\vec{v} = \{M_1^2,M_2^2\}$ is required and the weight is obtained by comparing Eqs.~\eqref{eq::simpledec2_nonlep} and \eqref{eq::evtgen_nonlep}:
\begin{align}
    w(M^2,\tilde{M}^2) = 4 M \tilde{M} \frac{|\tilde{V}_{B\rightarrow R_1 R_2}(M^2,\tilde{M}^2) |^2}{\hat{V}(M^2)\bar{V}(\tilde{M}^2)}~.
\end{align}
While the expression itself is less complex and does not involve non-trivial integrals, the weights are singular at $M^2 = (M_B - M_{R_2})^2$ and $\tilde{M}^2 = (M_B - M_{R_1})^2$. Sample weight distributions for three different $D^0$ decay modes are shown in Fig.~\ref{fig:DKstarRho_weights}.
All three cases exhibit a tail of large weights, in particular for $D^0\rightarrow K^{\ast 0}\bar{K}^{\ast 0}$ and $D^0\rightarrow \rho^+\rho^-$ decays. The two symmetric cases feature longer tails, since both resonances are either heavier ($K^{\ast 0}\bar{K}^{\ast 0}$) or wider ($\rho^+\rho^-$) compared to the asymmetric case. Consequently, more events are generated in the problematic region.
In practice, the large weights will lead to fluctuations and a signficant statistical uncertainty in the reweighted samples near the fake phase-space boundaries.

\section{Conclusion}
\label{sec:conclusion}
In this work we discussed two phenomenologically relevant hadron decay topologies that are simulated incorrectly by the commonly used event generator EvtGen: semileptonic decays into resonances, as well as hadron decays into two resonances. In both cases, the resonance invariant mass is generated by neglecting or oversimplifying phase-space factors and form factors. We demonstrated that the kinematic spectra produced by EvtGen show unphysical features and the shape distortions are sizeable, in particular for broad resonances and resonances near the phase-space boundary. For both cases we presented a method to derive correction weights for simulated events, allowing ongoing experimental analyses to correct the generated distributions.

In addition to the two decay topologies discussed here, phase-space factors are also neglected in decays featuring resonance to resonance decays, as well as decays with a sub-threshold resonance or three-body decays of a resonance. 
In the following, we summarize the known cases with missing factors:
\begin{enumerate}
    \item Semileptonic decays into a resonance, as discussed in Sec.~\ref{sec:sampling}. In this case form factors and phase-space factors relevant to the resonance production are neglected.
    \item Nonleptonic decays into two resonances, as discussed in Sec.~\ref{sec::nonlep}. For these decays, the phase-space factor at resonance production is evaluated incorrectly.
    \item Resonance decays into resonances, as described in Ref.~\cite{LHCb:2022jez}. Here, the phase-space factor of the decay resonance to resonance decay is evaluated at the nominal Breit-Wigner mass of the first resonance.
    \item Resonance decays into three or more final state particles as well as all decays in which the resonance mass lies below the production threshold of the final state. In these cases no phase-space factor is taken into account for the resonance decay. Furthermore, only a nonrelativistic Breit-Wigner function is used as propagator. This behaviour is documented in the documentation (\verb|doc/guide.ps|) accompanying the EvtGen source code.
\end{enumerate}

While reweighting of specific decays can serve as a short-term solution for affected analyses, it is clearly not optimal in the long term.
Going forward a new phase-space sampling algorithm needs to be implemented in EvtGen. Experimental collaborations focusing on precision flavour and hadron physics should consider the use of additional event generators, such as Sherpa or Herwig, to cross-check their main simulations. Comparative studies of different event generators and standalone predictions for selected benchmark processes, similar to efforts in Higgs physics \cite{Buckley:2021gfw,Chen:2025whf} or low-energy hadronic cross-sections in $e^+e^-$ collisions \cite{Aliberti:2024fpq}, are in order.

\section*{Acknowledgements}
We thank Stefan H{\"o}che for helpful comments on the manuscript, Tommy Martinov for suggesting to include the $w$ spectrum in Fig.~\ref{fig:BD1pmunu_q2El_comp} and John Back for clarifications regarding the EvtGen code.

\paragraph{Funding information}
FH is supported in part by the Swiss National Science Foundation (SNF) under contract 200021-212729.
RvT is supported by the German Research Foundation (DFG) Walter-Benjamin Grant No.~545582477.

\appendix
\section{Discussion of the EvtGen code}\label{app::code}
In this appendix we discuss the segment of the EvtGen code leading to the distortions discussed in this work. We refer to line numbers and file names in version 02.02.03 of EvtGen.

The incorrect or neglected phase-space factors can both be traced back to the same source file \verb|EvtRelBreitWignerBarrierFact.cpp|. Here, birth factors, comprising all phase-space and form factors, are initialized in \verb|EvtRelBreitWignerBarrierFact::getRandMass|, which is called once for every resonance in the decay. In lines 209 to 235, the mass of the sister resonance and the decaying particle in the function is set by:
\begin{verbatim}
double massOthD = -10.;
double massParent = -10.;
int birthl = -10;
if ( othDaugId ) {
    EvtSpinType::spintype spinOth = EvtPDL::getSpinType( *othDaugId );
    EvtSpinType::spintype spinPar = EvtPDL::getSpinType( *parId );

    int tt1 = EvtSpinType::getSpin2( spinOth );
    int tt2 = EvtSpinType::getSpin2( spinPar );
    int tt3 = EvtSpinType::getSpin2( _spin );

    //figure the min and max allowwed "spins" for the daughters state
    if ( ( tt1 <= 4 ) && ( tt2 <= 4 ) ) {
        birthl = std::max( tt3 - tt2 - tt1,
                    std::max( tt2 - tt3 - tt1, tt1 - tt3 - tt2 ) );
        if ( birthl < 0 )
            birthl = 0;

        massOthD = EvtPDL::getMeanMass( *othDaugId );
        massParent = EvtPDL::getMeanMass( *parId );
    }

    // allow user to override
    for ( size_t vC = 0; vC < _userSetBirthPW.size(); vC++ ) {
        if ( *othDaugId == _userSetBirthOthD[vC] &&
            *parId == _userSetBirthPar[vC] ) {
            birthl = 2 * _userSetBirthPW[vC];
        }
    }
}
\end{verbatim}
Most of the code deals with high spins that are not supported by EvtGen, as well as user orverrides. However, the \verb|if| condition is only triggered in the case where there is exactly one sister particle to the resonance and, consequently, \verb|othDaugId| is set. Otherwise the negative default values of \verb|massParent| and \verb|massOthD| remain. Furthermore, the function \verb|getMeanMass| selects the nominal Breit-Wigner mass of a given particle, not the invariant mass of its decay products.

The birth factor is later constructed in the code block between lines 254 and 262:
\begin{verbatim}
if ( massParent > -1. ) {
    if ( _includeBirthFact ) {
        EvtTwoBodyVertex vb( _mass, massOthD, massParent, birthl / 2 );
        vb.set_f( _blattBirth );
        amp.setBirthVtx( vb );
        amp.addBirthFact();
        amp.addBirthFactFF();
    }
}
\end{verbatim}
Consequently, no birth factor is included for semileptonic decays. In addition, the birth vertex is initialized with the nominal, not the invariant mass, of the sister particle, leading to the pattern in Eq.~\eqref{eq::evtgen_nonlep}. Since \verb|massParent| is set to the nominal mass of the parent particle, this part of the code also leads to the issue observed in resonance to resonance decays in Ref.~\cite{LHCb:2022jez}.

Finally, the arbitrary number of $10000$ accept/reject tries are performed in the function \verb|EvtDecayAmp::makeDecay| defined in \verb|EvtDecayAmp.cpp|.

\bibliography{bibliography.bib}

\end{document}